\documentclass{SCIS2021}

\usepackage{graphicx} 
\usepackage{epstopdf}
\usepackage{multirow}
\usepackage{makecell}
\usepackage{hyperref}

\begin{document}
\ArticleType{REVIEW}
\Year{2020}
\Month{}
\Vol{}
\No{}
\DOI{}
\ArtNo{}
\ReceiveDate{}
\ReviseDate{}
\AcceptDate{}
\OnlineDate{}

\title{A Survey on the Network Models applied in the Industrial Network Optimization}{A Survey on the Network Models applied in the Industrial Network Optimization }

\author[1]{Chao Dong}{}
\author[1]{XiaoXiong Xiong}{}
\author[1]{Qiulin Xue}{}
\author[2]{Zhengzhen Zhang}{}
\author[1]{Kai Niu}{{niukai@bupt.edu.cn}}
\author[1]{Ping Zhang}{}

\AuthorMark{Author A}

\AuthorCitation{Chao Dong, Xiaoxiong Xiong, Qiulin Xue, Zhengzhen Zhang, Kai Niu, Ping Zhang}


\address[1]{The Key Laboratory of Universal Wireless Communications, Ministry of Education, \\Beijing University of Posts and Telecommunications, Beijing {\rm 100876}, China}
\address[2]{The Smart City School of Beijing Union University, Beijing {\rm 100024}, China}

\abstract{Network architecture design is very important for the optimization of industrial networks. The type of network architecture can be divided into small-scale network and large-scale network according to its scale. Graph theory is an efficient mathematical tool for network topology modeling. For small-scale networks, its structure often has regular topology. For large-scale networks, the existing research mainly focuses on the random characteristics of network nodes and edges. Recently, popular models include random networks, small-world networks and scale-free networks. Starting from the scale of network, this survey summarizes and analyzes the network modeling methods based on graph theory and the practical application in industrial scenarios. Furthermore, this survey proposes a novel network performance metric - system entropy. From the perspective of mathematical properties, the analysis of its non-negativity, monotonicity and concave-convexity is given.  The advantage of system entropy is that it can cover the existing regular network, random network, small-world network and scale-free network, and has strong generality. The simulation results show that this metric can realize the comparison of various industrial networks under different models.}

\keywords{Industrial network, Small-scale network, Large-scale network, Graph theory, System entropy}

\maketitle

\section{Introduction}

Network architecture models play a very important role in the study of industrial network performance. The network can be classified according to its scale, for example, it can be divided into large-scale network and small-scale network. In order to better describe the performance of the network, graph theory \cite{add_1,add_2,add_3,add_4,add_5,bollobasModernGraphTheory2013,bondyGraphTheoryApplications1976,westIntroductionGraphTheory2001,vansteenGraphTheoryComplex2010}  is theoretically introduced as an analysis tool. Various metrics in graph theory, such as node degree, adjacency matrix, etc., are used to analyze the performance of the network.

For small-scale networks, the topology of the network has strong regularity, such as ring network, star network, etc. This type of network can also find corresponding instances in real industrial network deployment, such as wireless sensor network \cite{Relay Node Deployment Based Small World Effect in Hierarchical Industrial Wireless Sensor Networks}\cite{small_world_27} and computer network \cite{Optimization of Wireless Sensor Networks inspired by Small World Phenomenon}.

For large-scale networks, the size of the matrix describing the network characteristics increases with the number of network nodes and edges. The original small-scale network analysis method is not applicable, so it is necessary to study the statistical characteristics of the large-scale network.
In large-scale networks, commonly used network models include random networks \cite{add_10,erdos59a,erdhosEvolutionRandomGraphs1960}, small-world networks \cite{add_11,add_12,add_13,wattsCollectiveDynamicsSmallWorld1998}, and scale-free networks \cite{add_6,add_7,add_8,add_9,barabasiEmergenceScalingRandom1999,fronczakAveragePathLength2004}.

Random network is an early classic model\cite{erdos59a}\cite{erdhosEvolutionRandomGraphs1960}, which introduces the probability of connection between nodes. At this point, the topology of the network is no longer fixed. In a random network, the average path length between nodes increases logarithmically with the node network size. This is a preliminary result describing the asymptotic properties of the network.

In the 1990s, small-world networks were proposed as a special random network \cite{wattsCollectiveDynamicsSmallWorld1998}. In this kind of network, the distance between any two nodes is small, and there is a relatively deterministic upper bound for the distance. This good connectivity relies on certain key nodes in the network. This small-world property can find corresponding applications in the real world. For example, both social networks and the Internet have the small-world property \cite{vega-redondoComplexSocialNetworks2007}.
In small-world networks, the clustering coefficient is an important performance metric. A high clustering coefficient indicates that the network has better connectivity, and the cost of establishing connections between nodes is low.

Different from the above two networks, the scale-free network \cite{barabasiEmergenceScalingRandom1999} is a network model in which the degree distribution of nodes changes inversely proportional to the degree value. Its node degree probability mass function is an inverse proportional function affected by a parameter $\gamma$ \cite{onnelaStructureTieStrengths2007}\cite{choromanskiScaleFreeGraphPreferential2013}. This means that the larger the node degree, the less likely it is to appear. A commonly used scale-free network model is called the BA model \cite{barabasiEmergenceScalingRandom1999}. In this model, both the average path length and the clustering coefficient have closed-form expressions. The analysis results of the BA model show that its connectivity properties are better than small-world networks \cite{fronczakAveragePathLength2004}. This kind of network also has corresponding mapping objects in the real network. For example, the paper citation network is a typical scale-free network.

Although the above network models have been applied in real-world network optimization, industrial networks cover a wider range. For example, air traffic network\cite{The Properties of Campus Road Traffic Networks}\cite{Analysis of air traffic network of China}, power network \cite{wattsCollectiveDynamicsSmallWorld1998}\cite{11111,Reliability assessment to large-scale power grid based on small-world topological model,Comparison Analysis of the Smass-World Topological Model of CHINESE And American Power Grids} occupies an important position in industrial network optimization. Networks between different industrial categories are quite different. For network model abstraction, a metric with high versatility is needed for comparison between different industrial networks. Therefore, in addition to summarizing the existing network models and industrial network applications, this paper further proposes a network metric called system entropy based on Renyi entropy. The metric can cover the regular network, random network, small-world network and scale-free network. And the paper gives the convexity analysis of system entropy for node degree distribution. Our simulation results show that the metric has strong generality and can perform horizontal comparisons of various industrial networks under different models.

This paper is organized as follows. In Section 2, commonly used definitions and network metrics in graph theory are given. And network models including random networks, small-world networks, and scale-free networks are summarized. In Section 3, the modeling methods in the existing industrial network is given and the analysis of typical industrial network characteristics including air traffic network and power network detailed. Afterwards, in Section 4, this paper discusses the properties of system entropy in detail, such as envelope non-negativity, monotonicity symmetry and convexity. The performance of system entropy is verified by simulation results in Section 5. In addition, the performance of system entropy of different networks is compared.

\section{Typical network topology theory}

\begin{figure}[!t]
	\centering
	\includegraphics[width=4.5in]{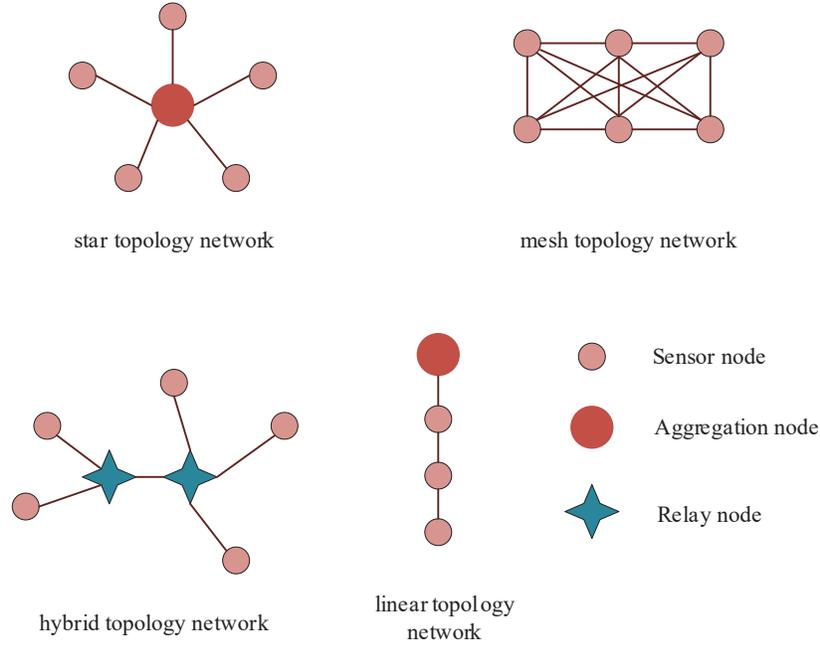}
	\caption{Small scale network topology.}
	\label{fig1}
\end{figure}

\begin{figure}[!t]
	\centering
	\includegraphics[width=5.4in]{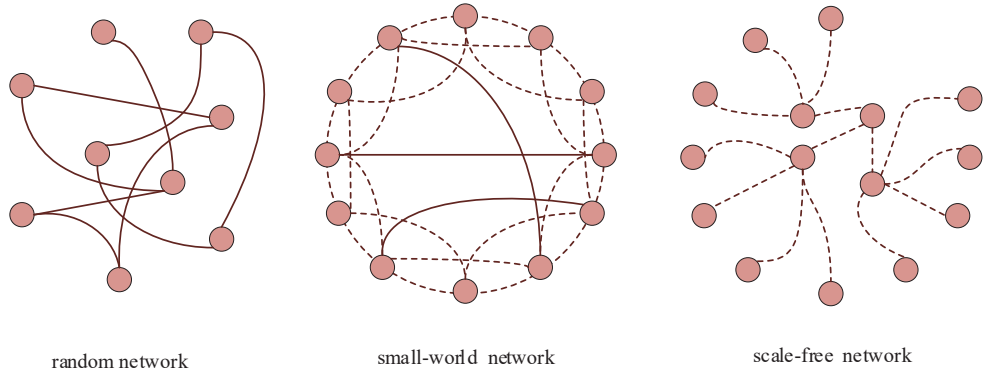}
	\caption{Large scale network topology.}
	\label{fig2}
\end{figure}

According to the size of the network, we classify the network into small-scale network and large-scale network. For small-scale networks, the network topology is relatively regular. For large-scale networks, the network topology is complex, and $complex$ $network$ is  needed to be introduced as tools for modeling and analysis.
Figure~\ref{fig1} shows the typical topology structure of  small-scale network, including star topology, mesh topology, hybrid topology, linear topology, etc. Furthermore, in Figure~\ref{fig2}, typical models commonly used in large-scale networks are given, including random networks, small-world networks, and scale-free networks.

Generally, networks are composed of multiple nodes. In addition, the associations between the nodes are represented by edges. Therefore, network topology can be abstracted as a graph, as shown in Figure~\ref{fig1} and Figure~\ref{fig2}. With the help of graph theory, the properties of network topologies can be systematically modeled and analyzed. In the following, section 2.1 gives common performance metrics in graph theory. Afterwards in section 2.2, the larger-scale network models, such as random networks, small-world networks and scale-free networks are briefly introduced.

\subsection{Basic Definitions and Performance Metrics in Graph Theory}

From the application point of view, graph theory \cite{bondyGraphTheoryApplications1976, westIntroductionGraphTheory2001, vansteenGraphTheoryComplex2010, bollobasModernGraphTheory2013} can be used for modeling and performance analysis of network topology.
Both nodes and connections in real networks can be represented by abstract concepts in graph theory. For example, in communication networks, base stations (BS), user equipment and sensors can all be modeled as nodes, and communication links between nodes can be modeled as edges. For the convenience of description, in subsection 2.1.1, we will give several basic definitions in graph theory. Afterwards, the performance metrics based on the basic definitions are introduced in subsection 2.1.2.

\subsubsection{Basic Definitions in Graph Theory}

\definition{A graph   consists of a collection $V$ of \textbf{\emph{vertices}} and a collection \textbf{\emph{edges}} $E$, which can be written as $G=\left( V,E \right)$. Each edge $e\in E$ joins two vertices. For example, $e=\left\langle u,v \right\rangle $ denotes that $e$ joins $u,v\in V$.}

$V\left( G \right)$ and $E\left( G \right)$ denote the set of vertices and edges associated with graph $G$, respectively.

\definition{For a graph $G$ and vertex $v\in V\left( G \right)$, the set of vertices (other than $v$) adjacent to $v$ $N\left( v \right)$  is the \textbf{\emph{neighbor}} set of $v$  , which can be defined as}

\begin{equation}
	N(v)\overset{\text{ }\!\!~\!\!\text{ def}}{\mathop{=}}\,\{w\in V(G)\mid v\ne w,\exists e\in E(G):e=\left\langle u,v \right\rangle \},
	\label{eq1}.
\end{equation}

\definition{The \textbf{\emph{degree}} of vertex $v$, $\delta \left( v \right)$ denotes the number of edges connected with  $v$.}

\definition{For a graph $G$ with $n$ vertices, a $n\times n$ diagonal matrix is used to represent the \textbf{\emph{degree matrix}} for $G$ which can be defined as }
\begin{equation}
	{{\bf{D}}_{i,j}}=\left\{ \begin{matrix}
		\delta \left( {{v}_{i}} \right) & if\text{ }i=j  \\
		0 & \text{otherwise}  \\
	\end{matrix} \right.
\end{equation}

The above concepts mainly describe the components of the graph. In order to be able to describe the characteristics of the graph as a whole, we need to introduce more complex matrices as measurement tools. Here, we mainly introduce the adjacency matrix and the Laplace matrix.

\definition{For a  graph $G$ with $n$ vertices, a square $n\times n$ matrix ${\bf{A}}$ is used to represent the \textbf{\emph{adjacency matrix}} of $G$  with entry ${{\bf{A}}_{i,j}}$ denoting the number of edges joining vertex ${{v}_{i}}$ and ${{v}_{j}}$. In an undirected graph, the adjacency matrix is a symmetric matrix.}

\definition{For a graph $G$ with $n$ vertices, its \emph{\textbf{Laplacian matrix}} ${\bf{L}}$ is defined as}
\begin{equation}
	{\bf{L}}={\bf{D}}-{\bf{A}}.
\end{equation}

The \textbf{\emph{symmetric normalized Laplacian matrix}} is defined as
\begin{equation}
	{{\bf{L}}^{\text{n}}}:={{\bf{D}}^{-\frac{1}{2}}}{\bf{L}}{{\bf{D}}^{-\frac{1}{2}}}={\bf{I}}-{{\bf{D}}^{-\frac{1}{2}}}{\bf{A}}{{\bf{D}}^{-\frac{1}{2}}} .
\end{equation}

In the following subsections, based on the above basic concepts, we introduce several key performance metrics for evaluating graph features, including \textbf{degree distribution}, \textbf{average path length and clustering coefficient}.

\subsubsection{Performance Metrics in Graph Theory}

\definition{  The probability distribution of vertex degree $\delta \left( v \right)$ is denoted by $p(\delta(v))$.}

The degree can reflect the characteristics of the graph. For example, those nodes with a high vertex degree often represent the key nodes in a network. In addition, degree distribution can be used to measure the information contained in the graph structure. If a communication network has an uneven degree distribution, i.e., a few vertices have relatively high degrees in comparison to others, these high-degree vertices can be seen as hubs. Once these hubs are removed, the original network may be split into unconnected sub-networks.

\definition{Consider a graph $G$ and $u,v\in V\left( G \right)$. $d\left( u,v \right)$ denotes the length of a shortest $\left( u,v \right)$-path.}

Distance statistics can be used to see to what extent two networks are different or not, but also to give an indication of the relative importance of each of the nodes in a network. And there are several relative metrics. The eccentricity of a vertex $u$ means how far the farthest vertex from $u$ is positioned in the network; the diameter denotes the maximal distance in a network.

\definition{Consider a connected graph $G$  with vertex set $V$, and  let $d\left( u \right)$ denote the average length of the shortest paths from vertex $u$ to any other vertex $v$ in $G$:}

\begin{equation}
	\bar{d}(u)\overset{\text{ }\!\!~\!\!\text{ def}}{\mathop{=}}\,\frac{1}{|V|-1}\underset{v\in V,v\ne u}{\mathop{\sum }}\,d(u,v).
	\label{eq2}
\end{equation}

The \textbf{\emph{average path length(APL)}} $d\left( G \right)$ is defined as

\begin{equation}
	\bar{d}(G)\overset{\text{ }\!\!~\!\!\text{ def}}{\mathop{=}}\,\frac{1}{|V|}\underset{u\in V}{\mathop{\sum }}\,\bar{d}(u)=\frac{1}{|V{{|}^{2}}-|V|}\underset{u,v\in V,u\ne v}{\mathop{\sum }}\,d(u,v).
	\label{eq3}
\end{equation}

One advantage of the above indicators is that they can adapt to the scalability of the network, because in a large-scale network, a complete description of the network based on the adjacency matrix or Laplacian matrix is accompanied by huge complexity, and statistical indicators (such as distribution or gradual characteristics) can ingeniously characterize the network, but for the complex industrial Internet, the traditional indicators are a bit one-sided. The focus of this article is to explore a new and appropriate indicator that can describe the characteristics of the industrial Internet. This indicator will be proposed in a later section. The following introduction to traditional indicators hopes to provide inspiration for new indicators.

One of the necessary operations for network analysis is taking a look at vertex degrees. There are several properties to examine through vertex degrees. For example, those nodes with a high vertex degree often represent the key players in a network. In addition, degree distribution can be used to derive information on the structure of a network. For example, if a network has an  extremely uneven degree distribution, (a few vertices have relatively high degrees in comparison to others), these high-degree vertices play the role of hubs, of which the removal may divide the network into several parts.

Besides vertex-degree distributions, distance statistics is another metric for network analysis.

In a real network like the Internet or the transportation network, a short average path length facilitates the quick transfer of information or goods and reduces costs. However, to measure the closeness of nodes in a group, the APL is not so effective.  Another often used metric for network analysis is the clustering coefficient. This coefficient means: for a given vertex $v$, to what extent the neighbors of $v$ are also neighbors of each other.

\begin{figure}[!t]
	\centering
	\includegraphics[width=4.5in]{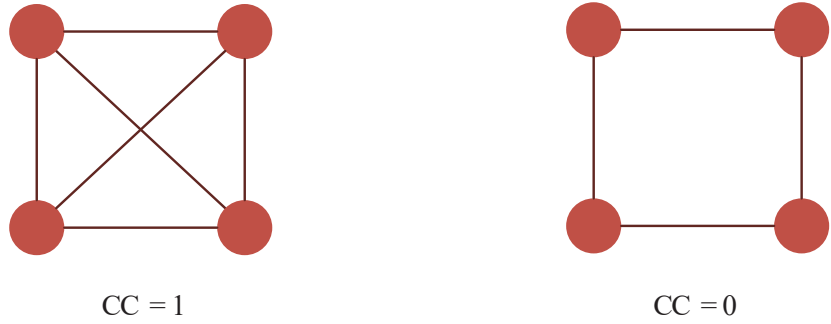}
	\caption{Example of clustering coefficient on an undirected graph.}
	\label{fig3}
\end{figure}

Here consider clustering coefficient introduced by Watts and Strogatz \cite{wattsCollectiveDynamicsSmallWorld1998}. From this perspective, the best clustering is that all neighbors are adjacent to each other. For example, Figure~\ref{fig3} shows a simplified representation of the clustering coefficient. The neighboring nodes of all nodes in the left figure are connected to each other, so the clustering coefficient is 1; the neighboring nodes of all nodes in the right figure are not connected, so the clustering coefficient is 0.
\definition{Consider a simple connected, undirected graph $G$ and vertex $v\in V(G)$ with neighbor set $N\left( v \right)$. Let ${{n}_{v}}=\left| N\left( v \right) \right|$and ${{m}_{v}}$ be the number of edges in the subgraph induced by $N\left( v \right)$, i.e., ${{m}_{v}}=|E(G[N(v)])|$. The clustering coefficient $cc\left( v \right)$ for vertex $v$ with degree $\delta \left( v \right)$ is defined as}
\begin{equation}
	cc(v)\overset{\text{ }\!\!~\!\!\text{ def}}{\mathop{=}}\,\left\{ \begin{matrix}
		{{m}_{v}}/\left( \begin{matrix}
			{{n}_{v}}  \\
			2  \\
		\end{matrix} \right)=\frac{2\cdot {{m}_{v}}}{{{n}_{j}}\left( {{n}_{v}}-1 \right)}\text{  if }\delta \text{(}v)>1  \\
		\text{ }\!\!~\!\!\text{ undefined                    otherwise}  \\
	\end{matrix} \right.
	\label{eq4}
\end{equation}

\definition{Consider a simple connected graph $G$. Let ${{V}^{*}}$ denote the set of vertices $\{v\in V(G)\mid \delta (v)>1\}$. The \textbf{\emph{average clustering coefficient(ACC)}} $CC\left( G \right)$ for $G$ is defined as}
\begin{equation}
	CC(G)\overset{\text{ }\!\!~\!\!\text{ def}}{\mathop{=}}\,\frac{1}{\left| {{V}^{*}} \right|}\underset{v\in {{V}^{*}}}{\mathop{\sum }}\,cc(v).
	\label{eq5}
\end{equation}

\subsection{Classic Network Models}

A classic network topology is regular networks, such as the topology of the base station in a cellular network. However, the application of  regular networks is relatively limited. The real-world network topology usually contains randomness. In order to describe the randomness, Paul Erd\"os and Alfr\'ed R\'enyi proposed the ER network model to describe pure random network \cite{erdos59a, erdhosEvolutionRandomGraphs1960}. On this basis, Watts \cite{wattsCollectiveDynamicsSmallWorld1998} and Barab\'asi \cite{barabasiEmergenceScalingRandom1999} successively introduced small-world networks and scale free networks through observation and research on real-world networks. The real-world network, such as transportation network, power network, industrial Internet, etc., is usually a topological structure between a regular network and a random network, and has similarities with the several classic network structures proposed above. As a basis for the description of the industrial Internet in the following section, this section introduces the four classic network topologies mentioned above: regular network, Erd\"os-R\'enyi random network, small world network and scale free network.

\subsubsection{Regular network}

Regular networks are more common in small networks, such as mesh network, grid network and so on. For example, regular network can often be seen in the performance analysis of wireless sensor networks (WSN), which are widely used for monitoring, detection and identification purpose by using different wireless sensor nodes. The deployment of sensor nodes in a WSN has been studied in \cite{mishraAnalysisDifferentGrid2015},which discussed the performance analysis of different grid types such as Triangular, Square, Pentagon, Hexagon, Heptagon and Octagon. Other than WSN, regular network also can be seen in cellular network as the classic hexagon network.
One of the main characteristics of a regular network is that the degrees of all nodes are basically equal, and the second is that the average path length is high because the distance between the remote node pairs is large.

\subsubsection{Random network}
Random networks have been introduced and studied for several decades. Paul Erd\"os and Alfr\'ed R\'enyi introduced what are now known as ``classical" random networks, or\textbf{\emph{ Erd\"os-R\'enyi networks}} \cite{erdos59a, erdhosEvolutionRandomGraphs1960}, in which  any two vertices are adjacent with some probability $p$.
The mean vertex degree of an $ER\left( n,p \right)$ graph is thus computed as
\begin{equation}
	\bar{\delta }\overset{\text{ }\!\!~\!\!\text{ def}}{\mathop{=}}\,\mathbb{E}[\delta ]\overset{\text{def}}{\mathop{=}}\,\underset{k=1}{\overset{n-1}{\mathop{\sum }}}\,k\cdot \mathbb{P}[\delta =k]\text{=}p\left( n-1 \right).
	\label{eq6}
\end{equation}

Consider the average path length for ER random graphs. First, Fronczak et al.  \cite{fronczakAveragePathLength2004} show that for (large) random graphs $H\in ER\left( n,p \right)$, the average path length can be estimated as
\begin{equation}
	\bar{d}(H)=\frac{\ln (n)-\gamma }{\ln (pn)}+0.5,
	\label{eq7}
\end{equation}
where $\gamma $ is the so-called Euler constant (which is approximately equal to 0.5772). Considering the derived mean vertex degree of ER random graphs, for large $n$, the average path length can be estimated as:
\begin{equation}
	\bar{d}(H)=\frac{\ln (n)-\gamma }{\ln (\bar{\delta })}+0.5,
	\label{eq8}
\end{equation}

For the clustering coeffcient, it is not difficult to see that for an $ER\left( n,p \right)$ random graph the expected value of the cluster coefficient is equal to $p$.

\subsubsection{Small world network}

A network is called a small-world network \cite{wattsCollectiveDynamicsSmallWorld1998} by analogy with the small-world phenomenon (popularly known as six degrees of separation). The small world hypothesis was first described by the Hungarian writer Frigyes Karinthy in 1929, and tested experimentally by Stanley Milgram (1967).

In 1998, Duncan J. Watts and Steven Strogatz published the first small-world network model, and the resulting graph $WS\left( n,k,p \right)$ is called a \textbf{\emph{Watts-Strogatz random graph}} \cite{wattsCollectiveDynamicsSmallWorld1998}. Their model proved that as long as a small number of links are added, a regular graph whose diameter is proportional to the size of the network can be transformed into a ``small world", where the average path length between any two vertices is very small, while the  clustering coefficient remains very large, which can be seen from the following theorem.

\theorem {For any Watts-Strogatz graph G from $WS\left( n,k,0 \right)$, the cluster coefficient for G is equal to}
\begin{equation}
	CC(G)=\frac{3}{4}\frac{(k-2)}{(k-1)}.
\end{equation}

What can be seen from the theorem is that the clustering coefficient of a $WS\left( n,k,0 \right)$ graph is independent of its size and that for large values of $k$ it is close to 3/4, which means that Watts-Strogatz graphs have high clustering coefficients.

As we all know, many networks exhibit the characteristics of small worlds, such as random graphs and scale-free networks \cite{Scale-free networks}. In addition, real-world networks such as the World Wide Web and Metabolic Networks also exhibit this characteristic.

In the industrial Internet, if the small-world feature is introduced, the distance and cost of information transmission can be reduced by reducing the distance between nodes, thereby effectively reducing energy consumption and delay. Therefore, studying the small world network is helpful to analyze and optimize the industrial Internet.


\subsubsection{Scale free network}

A scale-free network is a network in which the distribution of vertex degrees follows a \textbf{power law}, at least asymptotically. That is, the probability $P\left( k \right)$ that an arbitrary node has degree $k$ can be written as
\begin{equation}
	P(k)\sim {{k}^{-\gamma }},
\end{equation}
where $\gamma $ is a parameter whose value is typically in the range $2<\gamma <3$,  although occasionally it may lie outside these bounds \cite{onnelaStructureTieStrengths2007, choromanskiScaleFreeGraphPreferential2013}.

ER random networks have been defined as graphs where there is a probability that two vertices are adjacent. Watts-Strogatz networks are constructed by changing a well-structured graph by probabilistically repositioning its current edges between different vertices. However, the Watts-Strogatz model is usually used to describe small-world phenomenon, which cannot fully describe the properties of real-world networks (such as communication networks). The work of Barab\'asi and his student Albert has led to extensive research on scale-free networks. They showed that real-world networks, such as literature citations, actor collaboration, etc., exhibit a structure in which the number of high nodes decreases exponentially \cite{barabasiEmergenceScalingRandom1999}.

As pointed out by Dorogovtsev et al. \cite{dorogovtsevEvolutionNetworksBiological2013} and Vega-Redondo \cite{vega-redondoComplexSocialNetworks2007}, scale-free graphs can be constructed only through a growth process combined with what is referred to as \textbf{preferential attachment}.

It was Barab\'asi and Albert \cite{barabasiEmergenceScalingRandom1999} who first designed a scale-free network construction procedure.
The construction process combines the growth of the network with the attachment of new nodes to existing nodes with certain preferences (that is, the probability of selecting vertex u is proportional to its degree). The result graph is called a \textbf{\emph{Barab\'asi-Albert random graph}} , also referred as a $BA\left( n,{{n}_{0}},m \right)$ graph.

We can also apply the network analysis tools to attain insight in the properties of scale-free networks.

As regards to  the average path lengths,  \cite{fronczakAveragePathLength2004} derive the following estimation of the average path length for a $BA\left( n,{{n}_{0}},m \right)$ random graph
\begin{equation}
	\bar{d}(BA)=\frac{\ln (n)-\ln (m/2)-1-\gamma }{\ln (\ln (n))+\ln (m/2)}+1.5,
\end{equation}
where $\gamma$ is the Euler constant, which we also came across when estimating the average path length for ER random graphs.
BA graphs tend to systematically have a relatively much lower average path length than ER random graphs which already have very low average path length. So same as the small world network model, the average distance between two vertices in a BA network is very small relative to a regular network such as a lattice graph.

The analytical formula of the clustering coefficient for scale-free networks have been given by scholars as
\begin{equation}
	cc = \frac{{{m^2}{{(m + 1)}^2}}}{{4(m - 1)}}\left[ {\ln \left( {\frac{{m + 1}}{m}} \right) - \frac{1}{{m + 1}}} \right]\frac{{{{[\ln t]}^2}}}{t},
\end{equation}
where $t=n-{n}_{0}$ denotes the netowrk is generated by $t$-th step from ${n}_{0}$ initial nodes. It can be seen from this formula that when the network scale is sufficiently large, the BA scale-free network does not have obvious aggregation characteristics.

From the above analysis, it can be seen that features such as degree distribution and cluster coefficients can reflect the characteristics of topology and have certain physical meanings, but they can only represent a local feature. To completely characterize the network characteristics, it is also necessary to fully and comprehensively pay attention to the network topology. Therefore, new technical indicators need to be introduced which can take into account the network delay, energy consumption and other parameters.

\section{Industrial network topology}

In recent years, with the promotion of 5G technology \cite{2add6}\cite{2add8} and the rapid development of  Internet of Things and the Industrial Internet \cite{2add7}, intelligent inter-connection and massive access have become the salient features of the novel industrial networks \cite{2add1,2add2,2add3,2add4,2add5}. As the network scale expands, the optimization of the network architecture is important, otherwise it will affect the network construction time and the load of the sink nodes and hub nodes.

Therefore, the rational planning and layout of the novel industrial Internet has attracted much attention from both academia and industry. A reasonable topology for a specific industrial network scenario should be well configured according to functions of the network, such as behavior of the data flow, power consumption and latency. With the expansion of network scale, scalability and robustness against malicious attacks have become important indicators for network design evaluation. The scalability of the network is concerned with whether the addition of new nodes can minimize the impact on network performance. In addition, robustness against malicious attacks is concerned with whether the addition of a large number of new nodes will lead to a sharp increase in the pressure on the hub nodes, thereby affecting the performance of the network.

Based on the topological types introduced in the previous sections, this section enumerates the industrial scenes in the real world. Through these examples, it discusses how to select the proper regular topological structures in the real industrial scenes according to their own requirement and structural characteristics.
The characteristics of the overall topology of the network are analyzed based system dynamics under the trend of the rapid growth of network scale. The following three subsections give the details of optimization method and industrial examples of regular network, small-world network and scale-free network, respectively. Finally, in the fourth subsection, the existing network optimization methods are systematically summarized and their shortcomings are pointed out.

\subsection{Regular Network}
\begin{figure}
	\centering
	\includegraphics[width=14cm]{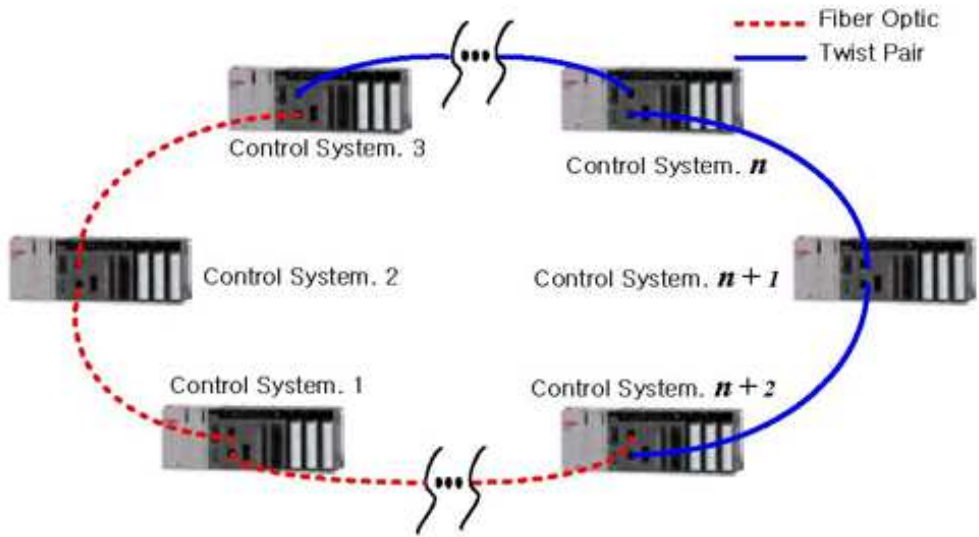}
	\caption{A ring topology of regular topology \cite{Ring Topology-based Redundancy Ethernet for Industrial Network}}
	\label{Ring}
\end{figure}

Regular topology is a widely used in industrial networks.
Different regular models are applied to describe their unique characteristics corresponding to real industrial networks.
The common network performance evaluation indicators include network diameter, average path length, clustering coefficient and fractal dimension.
In order to achieve the goal of network optimization, it is very important to choose a suitable topology.
Typical regular topology models  include linear topology, mesh topology, tree topology, cluster topology, cluster tree topology, ring topology, star topology and bus topology.
Specific cases about the application of the above topologies are introduced in the following.
In practical applications, industrial networks can use either a single-type topology or multi-type topology according to requirement.

\subsubsection{Single-type topology}
A reasonable network topology should well adapt to the real application scenarios.
The purpose of using a single-type topology is to maximize its benefits.
Three application instances of single type topologies are listed as follows.

WSN for the oil and gas industry are deployed using a linear topology that keeps these sensors at a distance along linear pipelines \cite{Performance Analysis of Linear Topology Wireless Sensor Network in Oil and Gas Industry}. This is due to the linear topology of oil and gas pipelines, and if the sensors are monitoring the pipe at a safe distance, the sensor network will also have a linear topology.

Redundant Ethernet topologies have the flexibility to change from ring topologies to wire (chain) topologies and vice versa. In redundant Ethernet, packets are guaranteed to reach their destination even if there is a link failure \cite{Ring Topology-based Redundancy Ethernet for Industrial Network}.
The star topology of common Ethernet is not flexible and reliable in industrial network due to its dependence on switches or hubs.
The redundant Ethernet system without switch and hub based on ring topology can solve this problem effectively.
A common problem in ring networks is that broadcast messages, if not blocked or consumed by some network sites, fall into an infinite loop of ring networks.
In \cite{Ring Topology-based Redundancy Ethernet for Industrial Network}, a logical polling mechanism is proposed to prevent unexpected packet storms in the network.

The topology of data center network (DCN) is mainly divided into switch centered network and server centered network.
In recent years, due to the rapid development of cloud computing applied in DCN, the traditional topology become unable to meet the requirements of DCN, and the introduction of FAT-Tree topology in DCN structure  \cite{A Survey of Network Topology of Data Center} solves the bandwidth bottleneck and single point failure problems of the traditional Tree structure.

\subsubsection{Multi-type topology}

Multiple topologies are used to combine the advantages of multiple topologies to adapt to more complex \cite{2add16} application scenarios. Two application instances of multi-type topologies are listed as follow.

In order to deal with the time-limited traffic in remote sensing/actuator control in production automation and monitoring in factory automation, a multi-hop cluster tree structure is applied to ensure the operation of sensor network in time-limited scenarios. This limited delay capability enables IEEE 802.15.4 cluster tree networks to support time-limited traffic \cite{A multichannel approach to avoid beacon collisions in IEEE 802.15.4 cluster-tree industrial networks}. In a cluster tree network, there are several levels of parent-child relationships among routers, and the lowest level determines the height of the tree. Because of this deterministic topology, the network can calculate the possible delay from the source to the coordinator.

In order to ensure good real-time property and network reliability, switched industrial Ethernet usually adopts a simple network topology with less than third order ring topology combined with tree topology, which is very popular in the industrial environment. Compared with the two components of this topology, the tree topology provides better real-time performance due to its lower forwarding delay, while the ring topology provides better network redundancy due to its fast recovery characteristics. The ring tree topology combines the structural advantages of the two topologies to construct a hybrid ring tree topology, using the ring topology in the upper layer to provide a reliable backbone, and using the tree topology in the lower layer to achieve a smaller forwarding delay  \cite{Graph partitioning strategy for the topology design of industrial network}.

\subsection{small-world Network}
As the network scale increases, so does the complexity of the network.
Complex networks have become one of the important research topics in many research fields.
Small-world \cite{Study on Small World Characteristics of In-Network Caching in Information-Centric Networks}  feature is one of the important features of complex networks, and its introduction into industrial network can give full play to its characteristics of small average path length, large clustering coefficient and scalability \cite{network architecture for data centers}\cite{PuLP: Scalable multi-objective multi-constraint partitioning for small-world networks}.

Small-world networks can be regarded as complex networks between regular networks and random networks \cite{A survey on the topology of wireless sensor networks based on small world network model}. Many studies \cite{The Impact of Heterogeneous Spreading Abilities of Network Ties on Information Spreading} \cite{Small but slow world how network topology and burstiness slow down spreading} have proved that using small-world networks can significantly reduce the energy consumption and improve the transmission efficiency of networks.
In practical applications of large-scale networks, a large number of end-to-end hops lead to a significant increase in delay. Therefore, the introduction of a small number of long-range connected edges can make a regular network or a random network into a small-world network. The small-world network feature reduces network latency \cite{Delay Optimized Small-World Networks}.

Industrial networks have inherent requirements for low latency \cite{Exploiting Small World Properties for Message Forwarding in Delay Tolerant Networks}\cite{A Fast Algorithm for Community Detection of Network Systems in Smart City} and high energy efficiency \cite{Relay Node Deployment Based Small World Effect in Hierarchical Industrial Wireless Sensor Networks}\cite{Multiobjective Small-World Optimization for Energy Saving in Grid Environments}. Nowadays, the small-world property has existed in many industrial networks, such as traffic networks, power networks and WSN networks.

\subsubsection{Traffic Network}

\begin{figure}
	\centering
	\includegraphics[width=14cm]{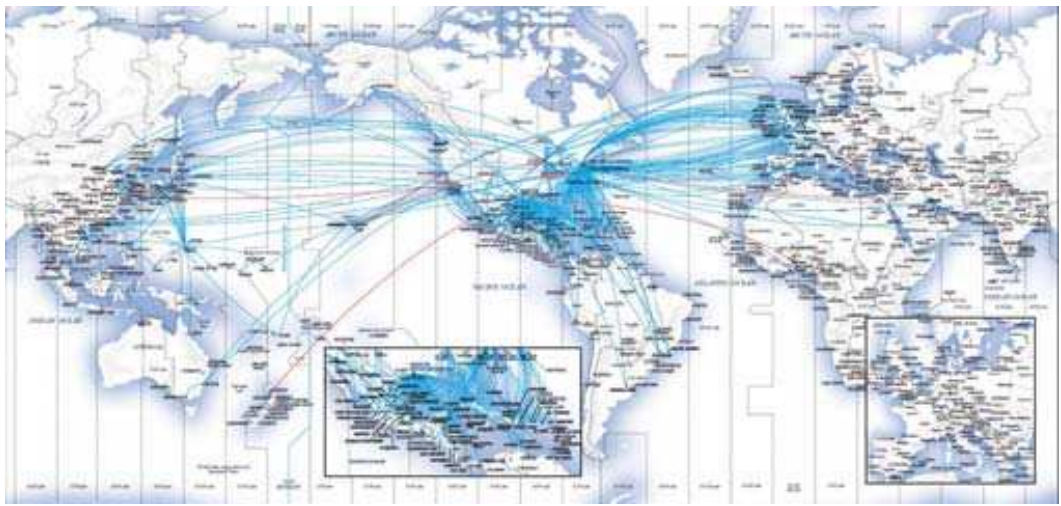}
	\caption{Topology of Global Aviation Network \cite{Reorganizing A New Generation Airline Network Based on An Ant-Colony Optimization-Inspired Small-World Network}}
	\label{ANC}
\end{figure}

Traffic network is a real network with small-world property \cite{2add17}\cite{2add18}. The main problem in traffic network is the delay caused by queuing, which is very similar to the congestion \cite{2add20} problem in communication network theory.

Traffic network tend to be inherently small-world. In  \cite{The Properties of Campus Road Traffic Networks}, five typical campus road traffic networks are considered,  including Huazhong University of Science and Technology, Wuhan University, Wuhan University of Technology, Central South University and Sun Yat-sen University.
Herein, the road intersection is seen as the node and  the road between intersections is seen as the network edge. Therefore, the above campus road network is studied based on the complex network theory \cite{2add19}, which provides the foundation and guidance for the traffic construction and management. Based on the complex network theory, \cite{The Properties of Campus Road Traffic Networks} calculates and analyzes the average degree, average clustering coefficient, average path length, network density and network tightness, etc, and the results show that the campus road traffic network has small-world behavior.

Another example of a small-world network in traffic is the air traffic system shown in figure.\ref{ANC}. With the increase of air traffic requirement, Chinese air traffic system (ATS), including China Airport Network (ANC) and China Airline Network (ARNC) \cite{Analysis of air traffic network of China} , faces great challenges such as delay growth, delay propagation and airspace congestion. To support such a huge passenger flow, small world attribute is indispensable. ANC network is a small-world (SW) network with short path length (5.05) and high clustering degree (0.7659). Given the growing demand for future air transport and the inherent capacity constraints, it is necessary to further improve network performance.

It is precisely because of the small-world nature of the air traffic network that the method of increasing the long-range connected edges can be applied to improve the performance of the network.
In accordance with this goal, the new generation of Boeing Airbus has the characteristics of lightweight fuselage and high energy conversion efficiency, and the direct flight mileage is greatly increased.
This attribute perfectly realizes the long-range edge in the small-world network, so it can be introduced into the planning of the new generation of aviation network to optimize the small-world attribute. Combining the aviation knowledge background with the characteristics of the new Airbus, \cite{Reorganizing A New Generation Airline Network Based on An Ant-Colony Optimization-Inspired Small-World Network} explores a small world network inspired by ant colony optimization, whose average path length is shorter than any traditional complex network. The generated  network architecture is called multi-star network, which can be used to achieve a shorter average distance and fewer transit times from any starting point on the earth to any destination.

\subsubsection{Power Network}
The power systems \cite{2add9,2add10,2add11,2add12,2add13,2add14} are also networks with small-world properties \cite{11111}. From the point of view of the system containing networks of different voltage levels, the power supply and distribution networks of middle and low voltage levels in cities are closely connected, while the transmission networks of high voltage levels are sparse. From the point of view of the transmission network of the same voltage stage, the network of each region is closely connected, while the network of different regions is sparsely connected. This indicates that the power system has the characteristics of small world network with high local clustering and poor global interconnection. Therefore, it is natural to think of the power system as a small-world network from a physical concept, for  example, East China Power Grid \cite{Reliability assessment to large-scale power grid based on small-world topological model}, West China Power Grid \cite{Collective dynamics of 'small-world' networks} and North China Power Grid \cite{Comparison Analysis of the Smass-World Topological Model of CHINESE And American Power Grids}.

However, the Small-world power network is vulnerable to cascading faults\cite{Comparison Analysis of the Smass-World Topological Model of CHINESE And American Power Grids} \cite{2add15}, so it also needs to be improved to accommodate the complex cascading faults. \cite{11111} makes two modifications to the original small-world network model: the line impedance not considered in the original model is considered as the weight of edges, and the plurality trend is introduced as the network flow in the small-world model. This improvement not only provides a new method for constructing small-world models for power network, but also provides a new measurement for network performance.

\begin{figure}
	\centering
	\includegraphics[width=12cm]{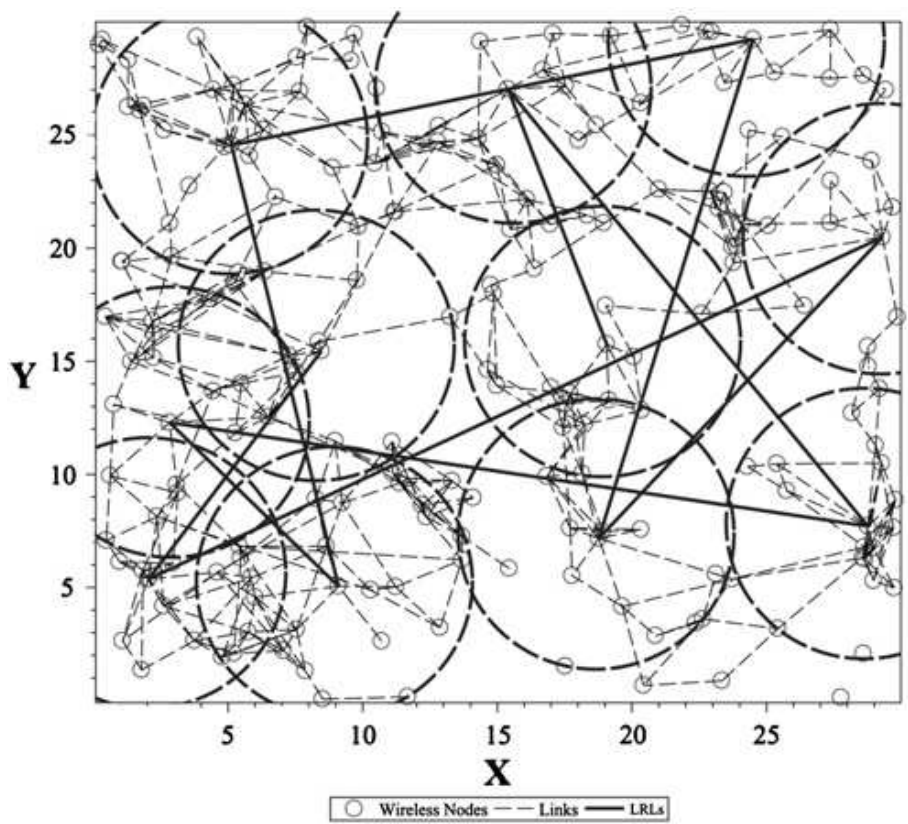}
	\caption{Wired and wireless hybrid structure \cite{Cost-effective design and evaluation of wireless sensor networks using topology-planning methods in small-world context}}
	\label{WSN}
\end{figure}

\subsubsection{Wireless Sensor Network}

WSN \cite{2add25}\cite{2add27} as a kind of complex network, has attracted worldwide attention \cite{2add21,2add22,2add23,2add24,2add28,2add31} for its superior performance in recent years. Sensor nodes are typically low-power and non-rechargeable in harsh outdoor conditions \cite{A survey on the topology of wireless sensor networks based on small world network model}. How to design a high energy efficiency network is an important challenge for WSN. Small-world networks have the characteristics of small average path length and the largest clustering coefficient, which can be used to optimize the energy \cite{small_world_27}, routing \cite{Optimization of Wireless Sensor Networks inspired by Small World Phenomenon}\cite{2add26,2add29,2add30} and topology \cite{Cooperative localization in small world wireless sensor networks}\cite{Joint Localization and Data Gathering Over a Small-World WSN With Optimal Data MULE Allocation} of wireless sensor networks.

It is well known that the small-world phenomenon can be used to reduce path length and energy consumption in WSN. However, the network built based on different small-world models will have subtle differences, which will affect the energy consumption of the above two optimization goals. Different small-world models will be selected to build the network framework. \cite{On the Analysis of Newman Watts and Kleinberg Small World Models in Wireless Sensor Networks} studied the effect comparison of two classical small-world models, Newman-Watts and Kleinberg, applied to WSN. It is shown that both models can create small-world sensor networks. Newman-Watts model has better results in path length, clustering coefficient and data communication delay. Kleinberg model can reduce more energy consumption in the process of data communication. This is because the Newman-Watts model does not take distance into account, and the created shortcuts can shorten the path length compared with the Kleinberg model. Consequently, the data communication latency is lower.

Unfortunately, in the above small-world WSN topology, sensor nodes consume most of the energy when sending and receiving packets over wireless long link. This often requires the introduction of wires to reduce the number of hops in the wireless network, thus reducing the energy consumption per node.
The optimization of energy efficiency of sensor networks aims to reduce the energy consumption of nodes and balance the energy consumption between nodes. Studies in \cite{small_world_27} have shown that adding a small number of wires into the WSN can not only reduce the number of hops of the sensor network, but also balance the energy consumption of the network, prolong the life of the network and improve the robustness of the network. It has been shown that the use of limited infrastructure (wired shortcuts, as figure.\ref{WSN}) can improve the energy efficiency of WSN. But over time, some nodes in the network will first run out of energy, which accelerates the energy depletion of others and makes the network a disconnected topology. Therefore, it is favored to  obtain a high clustering coefficient and a small path length through the small-world model. These property can not only balance the energy consumption but also keep the network connected after some nodes run out of power. Thus, the network lifetime and the robustness of the network are elevated.

In order to review the application of small world network in WSN more comprehensively, the performance indicators concerned by WSN in small world model mainly include the following items \cite{Optimization of Wireless Sensor Networks inspired by Small World Phenomenon}
\begin{itemize}
	\item [1)]
	Network dynamics. Whether the small-world WSN network can adapt to the mobility of nodes in the network.
	\item [2)]
	Node deployment. Whether small-world WSN can generate routes for ad-hoc networks efficiently and adaptively.
	\item [3)]
	Energy considerations. Small-world WSN needs to decide whether to use one hop or multiple hops according to geographical factors and energy consumption factors.
	\item [4)]
	The function of nodes. Whether the small-world WSN can accommodate nodes with multiple functions and coexist in the network.
	\item [5)]
	Control package overhead. Synchronize and other control operations with a minimum number of control packs to minimize energy consumption.
	\item [6)]
	Quality of service (QoS). Whether the small-world WSN need energy-aware routing protocol to achieve the compromise between QoS and node lifetime.
\end{itemize}

Considering the various optimization objectives mentioned above, the classical small-world network can no longer satisfy such different directions. Therefore, the enhanced version of small-world network is used for optimization in most studies as follow.

Firstly, the most direct enhancement is to improve the generation logic of the traditional small-world network. Wireless networks are not inherently small worlds, and it is not easy or cost-effective to manually create networks with this feature using existing technology. In other words, the traditional blind rewiring techniques designed to enhance networks with these characteristics suffer from inefficiency and saturation behavior. \cite{Cost-effective design and evaluation of wireless sensor networks using topology-planning methods in small-world context} proposes a topology planning approach that effectively adds the small-world nature of the network by using expensive long-distance transmission facilities. Essentially, this approach is enhanced from the dynamics of network evolution. To demonstrate the ability of these methods to deal with real situations, \cite{Cost-effective design and evaluation of wireless sensor networks using topology-planning methods in small-world context} tested networks with `clustering coefficients' and `diameters' falling into different ranges. The results show that the combination of these technologies reduces the `diameter' of the network by nearly 50\% and the `average path length' by 47\%. This reduces operation overhead by 67\% compared to traditional blind rewiring.

More applications of the enhanced small-world model are listed in figure.\ref{1}, and will not be discussed here.

\subsection{Scale-Free Network}
Recent research on the topology of the industrial Internet has shown that many networks are controlled by a relatively small number of hub nodes that are connected to many other nodes. For example, the World Wide Web is a network of virtual Web pages connected by hyperlinks that are connected by a Uniform Resource Locator (URL) for navigating from one Web page to another. Through the study of the network of biotech industry alliances in the United States, it is found that companies such as Genzyme, Chiron and Genentech are the clear center of the network, and they have a disproportionately large number of cooperative relationships compared with other companies.

The hub nodes are born due to the system dynamics principle of scale-free network. The addition of new nodes does not affect the internal structure of the network, and the connection mechanism of new nodes leads to the scale-free topology being defined as a power-law distribution. The emergence of scale-free structures and power-law degree distributions can be attributed to two mechanisms -- growth and preferred attachment \cite{Community Detection Algorithm Based on the Scale Free Property of Networks} -- that as new nodes appear, they tend to connect to the nodes which are more linked, and over time, these popular hubs gain more connections than their less connected neighbors. This "getting richer" process often favors early nodes, which are more likely to end up as hubs. A growing network with preferred connections will indeed become scale-free, and its node distribution follows a power law.

In scale-free networks, as \cite{Scale-free networks} introduced, the existence of hub nodes is a coin with two sides. This particular topology has distinct advantage and disadvantage, i.e., efficiency and vulnerability. The introduction of this subsection includes the following two parts. The first part introduces the benefits of attaching scale-free property to industrial networks, and the second part introduces the vulnerability of scale-free industrial networks.

\subsubsection{Advantages of Scale-Free industry Network} \label{Advantages of Scale-Free industry Network}

Scale-free feature is conducive to the efficient operation of industrial Internet. The scale-free architecture can improve the end-to-end performance, which is due to the reduction of the average path length and the reduction of the traffic intensity on the bottleneck link. Further, this phenomenon is entirely due to these hubs in the network, and associated with this feature, end-to-end data transmission requires only a handful of transferrings. From the above description, it is not difficult to find that scale-free networks have small-world property to some extent due to the existence of these hubs \cite{Scale-free networks}.

All kinds of original scale-free industrial networks embody the characteristics of small world. On the World Wide Web, with more than 3 billion documents, it typically takes only 19 clicks between pages \cite{Scale-free networks}. For the 800 million nodes of the WWW in 1999, the typical shortest path between two randomly selected pages is also only around 19 \cite{The Architecture of Complexity}.

Inspired by the above properties of scale-free networks, the researchers try to introduce scale-free features into the topology of existing networks to improve its performance.
For example, \cite{On the Impact of Scale-Free Structure on End-to-End TCP Performance} analyzes the end-to-end performance (send rate, probability of packet loss, and round-trip time) of TCP streams over a scale-free network, which benefits from decreasing the maximum of link betweennesses. For the same reason, \cite{Scale-free topology for large-scale wireless sensor networks} proposes an Arbitrary Weight based Scale-Free topology control algorithm (AWSF) which introduces the scale-free characteristics of complex networks into the topology of WSN to reduce transmission delays. In addition, \cite{Generating Scale-Free Topology for Wireless Neighborhood Area Networks in Smart Grid} and \cite{Research on Scale-Free Network User-Side Big Data Balanced Partition Strategy} introduce the scale-free characteristics into smart grid and Big Data, respectively, and both achieve greater efficiency.

\subsubsection{Vulnerability to malicious attacks}

As mentioned in \ref{Advantages of Scale-Free industry Network}, the performance advantage of scale-free network comes from the fact that the existence of hubs greatly reduces the average distance between nodes in the whole network. However, reliance on hubs has a serious drawback: vulnerability to malicious attack \cite{Scale-Free Networks: Characteristics of the Time-Variant Robustness and Vulnerability}. In a series of studies, it was found that simply removing a few key hubs from the Internet would split the entire system into isolated sets of routers. Eliminating 5 to 15 percent \cite{Scale-free networks} of the hub nodes in the face of a specific attack could crash a system. In addition, the hubs in WSN will have greater forwarding burden and energy consumption than nodes with lower degree. Therefore, it is particularly important to design an efficient and robust WSN.

\cite{Onion-like network topology enhances robustness against malicious attacks} proposed a structure of which the degree of nodes decreases from the center of the network to the boundary. It was named as the onion structure. This gradually decentralized structure has been shown to be more robust to specific attacks, which provides a crucial idea for industrial Internet to withstand specific attacks.

In order to further improve the performance of scale-free networks, some novel methods are proposed. The scheme energy-aware low potential-degree common neighbor (ELDCN) \cite{Toward Energy-Efficient and Robust Large-Scale WSNs: A Scale-Free Network Approach} provides another design idea - As the network expands, ELDCN avoids new nodes establishing links to hub-nodes with high potential connectivity, which takes both energy-efficient and robust into account. The Improved Scale-Free Network (ISFN) \cite{Robustness Optimization of Scale-Free IoT Networks} technique reduces the vulnerability of the network. ISFN consists of the edge degree-based and eccentricity based swap operations which kept the degree of the nodes unchanged (this means that the network remain scale-free). A more flexible tunable coefficients are introduced in \cite{Scale-free model for wireless sensor networks} to adjust the structure of the networks to achieve balance between connectivity and energy consumption. This way of design can better match the real scenarios with requirements of concrete compromise.

\begin{figure}
	\centering
	\includegraphics[width=14cm]{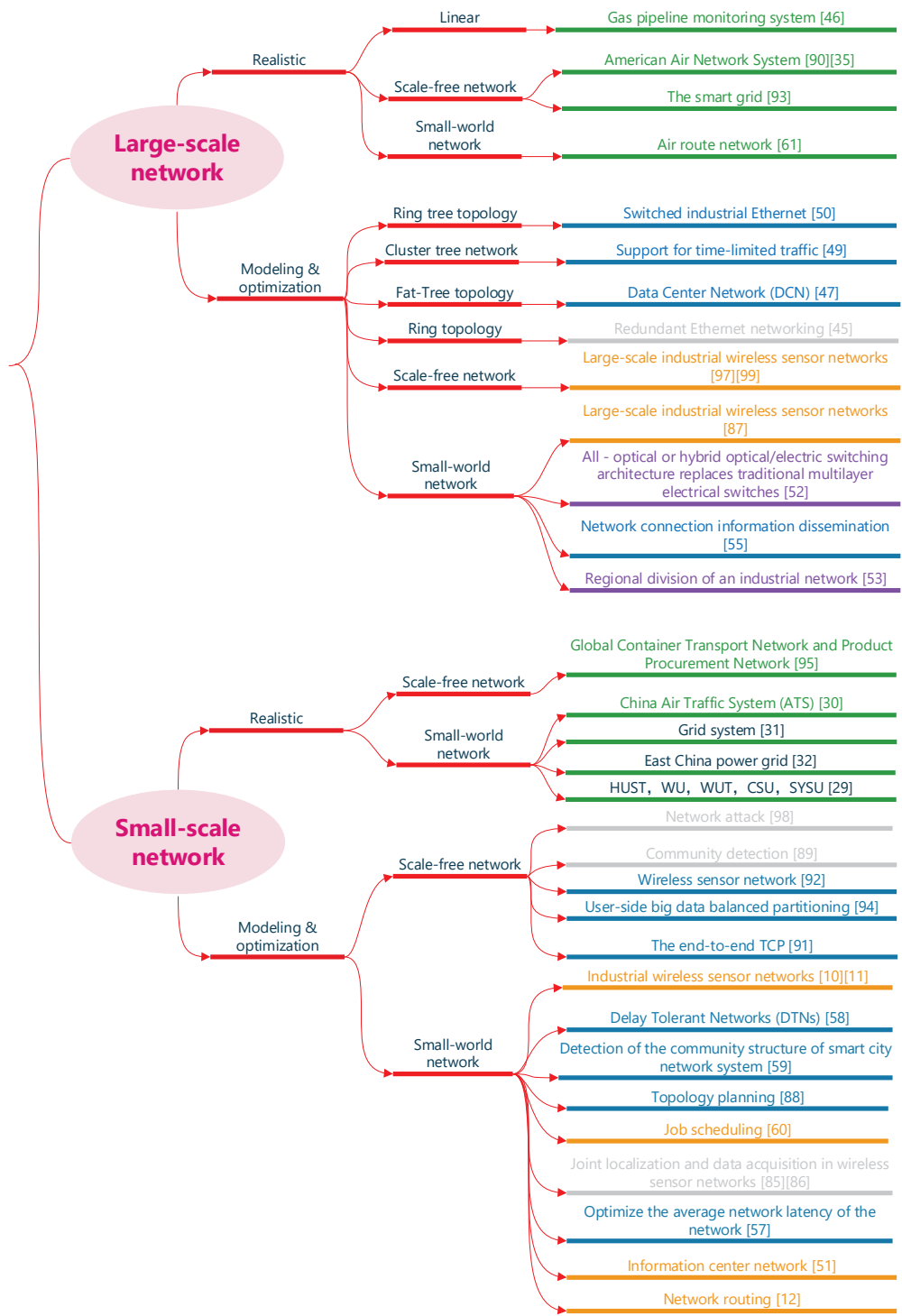}
	\caption{Summary of industrial Internet scenarios \& topologies}
	\label{1}
\end{figure}

\subsection{Summary and the Lack of Systematic Measurement}

The first three sections have systematically summarized the research status of the industrial Internet, using a variety of topology structure for a certain or several performance criteria of optimization, basically the following, power consumption, delay, ductility, robustness.
Figure.\ref{1} summarizes the industrial interconnection network topology mentioned above. The green bottom line represents an example of the industrial Internet. Other than green bottom line, the different color bottom lines represent the different optimization goals that the article focuses on. The blue, gray, yellow and purple of the bottom line represent delay, robustness, power consumption and scalability, respectively. However, the analysis also found that the relationship between these performance criteria is mostly negative correlation. Through comprehensive consideration of the performance criteria of the optimized network, it is not difficult to find the following phenomenon.

The above description can lead to the conclusion that some method can achieve balance loading and energy saving. However, some high degree nodes are naturally suitable as hub nodes. Thus under such method, in order to realize the loading balance, many paths in routing list will be forced to extend their hops and many shortcuts through hubs will be sacrificed. Similarly, although the decentralized operation of scale-free network reduces the power consumption of the hub, it will also lead to the whole network being forced to schedule more nodes and consume more energy for transmitting.

For another example, the advantage of small-world network lies in the existence of long-range edges. The introduction of small-world attribute to random network can make the existence of shortcuts in the network, and these shortcuts can speed up the data packets transmission in the network. However, there is a kind of cost problem with this approach. When the above connection demand exceeds the wireless coverage ability of industrial nodes, expensive wired equipment is often needed to realize the long-range coverage of the industrial network. In the process of introducing small-world attribute to existing network, due to the lack of systematic guidance for this scenario, the introduction of wired connection can only rely on blind "edge breaking and reconnection" to achieve small-world network topology. This kind of uncertainty is not desirable in the face of expensive wired equipment costs.

From these examples, it is not difficult to find that network relaying efficiency and node relaying pressure are inversely proportional to each other. Under the existing optimization ideas, the optimization of one of the two needs to sacrifice the performance of the other to some extent. If the modern industrial network seeks to break through this bottleneck, it needs to optimize the two directions jointly. However, there seems to be no such performance criterion to evaluate the combination of the two performance criteria, and no such tool to complete the joint optimization.

According to the above analysis, it can be found that the improvement of node relay performance can lead to the increase of node energy consumption.
In order to more fully optimize the network, it requires a method of global analysis and joint optimization, which can be used to integrate most of the industrial Internet performance criteria into the unified system for discussion.


\section{System Entropy theory}
The previous chapter gives a systematic overview of the current situation of modern industrial network based on the performance criteria optimized by various existing articles, from which it can be concluded that the cost of optimization often needs to sacrifice the performance of other performance criteria. This leads to the thinking that, since the evaluation performance criteria in the previous chapter are basically related to the topology, can a new performance criterion be proposed to evaluate the network topology, so as to directly or indirectly reflect the system performance of the network?

Combined with the above research status, the concept of ordered criterion and system entropy is creatively proposed, which leads to a network measurement theory based on network topology. The ordered criterion based on system entropy is more abstract than the existing specific criteria such as power consumption and delay, which can be used to measure the overall performance of the network from the perspective of network architecture, and guide the layout of industrial network and the topology of the connections between industrial network nodes.

In this paper, Renyi entropy of 0.5 order is used as the system entropy expression to describe the ordered criterion of network topology as follow,
\begin{equation}
{{H}_{\alpha }}\left( {{\mathbf{L}}^{n}} \right)=\frac{1}{1-\alpha }{{\log }_{2}}\left( tr\left( {{\left( {{\mathbf{L}}^{n}} \right)}^{\alpha }} \right) \right).
\label{symH}
\end{equation}

As is known to all, the parameter $\alpha$ of Renyi entropy can be any value greater than 0. However, in eq.(\ref{symH}), $\alpha=0.5$ is the parameter of Renyi entropy, it will guarantee a positive entropy as below. And the matrix ${\mathbf{L}}^{n}$ is normalized Laplacian matrix. In this paper, the symmetric Laplace matrix is considered to be unweighted, which means that it is only used to describe the topology of the network. In order to prove that it has similar physical meaning with general entropy, several properties of system entropy are derived in this paper, and the convexity, extremality, symmetry and nonnegativity of system entropy will be discussed in the next four sections.

\subsection{Convex Property and Monotonicity}

In this subsection, the convex property and monotonicity of the system entropy will be introduced. The monotonicity of system entropy is based on convexity. Because with the convexity theorem of entropy function, the monotonicity of the function is obvious. According to the expression of convexity and the expression of system entropy, the convexity of the system entropy can be described by the following expression,

\begin{equation}
	{{H}_{\alpha }}\left( \beta \mathbf{L}_{1}^{n}+\left( 1-\beta  \right)\mathbf{L}_{2}^{n} \right)\ge \beta {{H}_{\alpha }}\left( \mathbf{L}_{1}^{n} \right)+\left( 1-\beta  \right){{H}_{\alpha }}\left( \mathbf{L}_{2}^{n} \right),
	\label{Convex}.
\end{equation}
inequality is constant when $\beta$ changes from 0 to 1. However, it is difficult to prove the convex property directly from the eq.(\ref{Convex}), because it involves the operation of inequalities of multiple parameters and matrices. We can think about proving starting with the system entropy function itself. Firstly, the function $\log \left( \cdot  \right)$ is convex over the domain $\left( 0,+\infty  \right)$, because,
\begin{equation}
\frac{{{\partial }^{2}}}{{{(\partial x)}^{2}}}\log (x)=\frac{-1}{{{x}^{2}}}<0,
\end{equation}

and it just need to prove that $tr\left( {{\left( {{\mathbf{L}}^{n}} \right)}^{\frac{1}{2}}} \right)$ is greater than zero. In order to prove this assumption, it first need the lemmas below to prove that normalized Laplacian matrix is real symmetric and non-negative definite matrix.

\lemma[]{}${\mathbf{L}}^{n}$ is a real symmetric matrix. Because it obeys the following expression,

\proof
\begin{equation}
{{\left( {{\mathbf{L}}^{n}} \right)}^{T}}={{\left( {{\mathbf{D}}^{-\frac{1}{2}}}\mathbf{L}{{\mathbf{D}}^{-\frac{1}{2}}} \right)}^{T}}={{\left( {{\mathbf{D}}^{-\frac{1}{2}}} \right)}^{T}}{{\left( \mathbf{L} \right)}^{T}}{{\left( {{\mathbf{D}}^{-\frac{1}{2}}} \right)}^{T}}={{\mathbf{D}}^{-\frac{1}{2}}}\mathbf{L}{{\mathbf{D}}^{-\frac{1}{2}}}={{\mathbf{L}}^{n}}
\end{equation}
$\hfill\blacksquare$
\lemma[]The normalized Laplace matrix is a nonnegative definite matrix.

\proof Because ${\mathbf{L}}^{n}$ is a real symmetric matrix, so it can be decomposed into
\begin{equation}
{{\mathbf{L}}^{n}}={{\left( {{\mathbf{V}}^{n}} \right)}^{T}}{{\mathbf{V}}^{n}}
\label{decomposed}.
\end{equation}

In combination with the eq.(\ref{decomposed}), there are no more than three cases of elements in matrix ${\mathbf{L}}^{n}$:

\begin{equation}
\left \{
\begin{aligned}
	& {{\left[ {{\left( {{\mathbf{V}}^{{n}}} \right)}^{T}}{{\mathbf{V}}^{{n}}} \right]}_{i,j}}=\frac{{{d}_{ii}}}{{{d}_{ii}}}\text{  },\text{  }i=j \\
	& {{\left[ {{\left( {{\mathbf{V}}^{{n}}} \right)}^{T}}{{\mathbf{V}}^{{n}}} \right]}_{i,j}}=-\sqrt{\frac{1}{{d}_{ii} {d}_{jj}}}\text{  },\text{  }i\ne j\text{ }and\text{ }there\text{ }exists\text{ }an\text{ }edge\text{ }between\text{ }i\text{ }and\text{ }j \\
	& {{\left[ {{\left( {{\mathbf{V}}^{n}} \right)}^{T}}{{\mathbf{V}}^{n}} \right]}_{i,j}}=0\text{  },\text{  }i\ne j\text{ }and\text{ }there\text{ }doesn't\text{ }exist\text{ }an\text{ }edge\text{ }between\text{ }i\text{ }and\text{ }j \\
\end{aligned} \right.
\end{equation}

It is noticed that the $\frac{{{d}_{ii}}}{{{d}_{ii}}}$ can be decomposed into $\sum\limits_{{{d}_{ii}}}{\frac{1}{{{d}_{ii}}}}$, combining the square formula, the quadratic type ${{\mathbf{x}}^{T}}{{\mathbf{L}}^{n}}\mathbf{x}$ can be disassembled as

\begin{equation}
{{\mathbf{x}}^{T}}{{\mathbf{L}}^{n}}\mathbf{x}=\sum\limits_{i,j\in E}{\left( \frac{1}{{{d}_{ii}}}x_{i}^{2}-\frac{2}{\sqrt{{{d}_{ii}}{{d}_{jj}}}}{{x}_{i}}{{x}_{j}}+\frac{1}{{{d}_{jj}}}x_{j}^{2} \right)}=\sum\limits_{i,j\in E}{{{\left( \frac{1}{\sqrt{{{d}_{ii}}}}{{x}_{i}}-\frac{1}{\sqrt{{{d}_{jj}}}}{{x}_{j}} \right)}^{2}}},
\label{post_define}
\end{equation}

where ${i,j\in E}$ means that there exists an edge between node $i$ and node $j$.

The eq.(\ref{post_define}) shows that if $\frac{1}{\sqrt{{{d}_{ii}}}}{{x}_{i}} \ne \frac{1}{\sqrt{{{d}_{jj}}}}{{x}_{j}}$, the quadratic type ${{\mathbf{x}}^{T}}{{\mathbf{L}}^{n}}\mathbf{x} > 0$  and if $\frac{1}{\sqrt{{{d}_{ii}}}}{{x}_{i}}=\frac{1}{\sqrt{{{d}_{jj}}}}{{x}_{j}}$ the quadratic type ${{\mathbf{x}}^{T}}{{\mathbf{L}}^{n}}\mathbf{x} = 0$. So the matrix ${\mathbf{L}}^{{n}}$ is a nonnegative definite matrix.

$\hfill\blacksquare$

Given these two important lemmas, we can then prove the nonnegativity of $tr\left( {{\left( {{\mathbf{L}}^{\text{n}}} \right)}^{\frac{1}{2}}} \right)$.

Firstly, if ${\mathbf{L}}^{\text{n}}$ is a real symmetric matrix, there must be an orthogonal matrix ${\mathbf{P}}$ to diagonalize it as follow,
\begin{equation}
{{\mathbf{L}}^{n}}=\mathbf{P}{{\mathbf{\Lambda }}^{n}}{{\mathbf{P}}^{-1}}
\end{equation}

Since ${\mathbf{L}}^{\text{n}}$ is a semi-definite matrix, the elements of diagonal matrix ${{\mathbf{\Lambda }}^{n}}$ are greater than zero.
And it's because of the orthogonality of the matrix ${\mathbf{P}}$, the matrix ${\mathbf{L}}^{\text{n}}$ can be further decomposed into

\begin{equation}
	{{\mathbf{L}}^{n}}=\mathbf{P}{{\left( {{\mathbf{\Lambda }}^{n}} \right)}^{\frac{1}{2}}}{{\mathbf{P}}^{\text{-1}}}\mathbf{P}{{\left( {{\mathbf{\Lambda }}^{n}} \right)}^{\frac{1}{2}}}{{\mathbf{P}}^{\text{-1}}}
\end{equation}

$tr\left( {{\left( {{\mathbf{\Lambda }}^{n}} \right)}^{\frac{1}{2}}} \right)\ge 0$ is clearly established, and then we can conclude, the matrix ${{\left( {{\mathbf{L}}^{n}} \right)}^{\frac{1}{2}}}$ is also a semidefinite matrix, and $tr\left( {{\left( {{\mathbf{L}}^{n}} \right)}^{\frac{1}{2}}} \right)\ge 0$.

The independent variable $tr\left( {{\left( {{\mathbf{L}}^{\text{n}}} \right)}^{\frac{1}{2}}} \right)$ is in the domain $\left( 0,+\infty  \right)$ of the convex function, the convexity of the system entropy function, ${{H}_{\alpha }}\left( \beta \mathbf{L}_{1}^{n}+\left( 1-\beta  \right)\mathbf{L}_{2}^{n} \right)\ge \beta {{H}_{\alpha }}\left( \mathbf{L}_{1}^{n} \right)+\left( 1-\beta  \right){{H}_{\alpha }}\left( \mathbf{L}_{2}^{n} \right)$, is proved.

As discussed at the beginning of this section, convexity can reflect the monotonicity of functions. The system entropy function is convex, so the monotonicity of the function is monotonically increasing. In particular, the system entropy function increases monotonically as $tr\left( {{\left( {{\mathbf{L}}^{n}} \right)}^{\alpha }} \right)$ increases.

\subsection{Symmetry}

Symmetry is an important characteristic of system entropy. Firstly, we will explain the physical meaning of symmetry in the systems entropy theory. Since different rows of the Laplace matrix represent different nodes, when a certain topology is mapped to a Laplace matrix, different Laplace matrices will be obtained because of different indexes of the definition of the nodes.

This property is the entropy-invariant property obtained by replacing the column and column vectors of a pair of nodes in the Laplace matrix when traversing all nodes. Without loss of generality, we begin with the case of replacing a pair of nodes. The substitution of a pair of nodes in the Laplace matrix requires the permutation matrix in the elementary matrix, marked as $\mathbf{M}$, to complete the operation. The form of this elementary matrix is as follows

\[\mathbf{ M } = \left[ \begin{matrix}
	1 & {} & {} & {} & {} & {} & {}  \\
	{} & \ddots  & {} & {} & {} & {} & {}  \\
	{} & {} & 0 & \cdots  & 1 & {} & {}  \\
	{} & {} & \vdots  & \ddots  & \vdots  & {} & {}  \\
	{} & {} & 1 & \cdots  & 0 & {} & {}  \\
	{} & {} & {} & {} & {} & \ddots  & {}  \\
	{} & {} & {} & {} & {} & {} & 1  \\
\end{matrix} \right],\]

and the permutation operation of the Laplace matrix is as follows

\begin{equation}
{{\left( \mathbf{L}_{2}^{n} \right)}^{\frac{1}{2}}}=\mathbf{M}{{\left( \mathbf{L}^{n} \right)}^{\frac{1}{2}}}\mathbf{M},
\end{equation}

where $\mathbf{L}_{2}^{n}$ is the Laplace matrix after the permutation operation. At the same time, from the lemma below, we noticed that the matrix $\mathbf{M}$ is also an orthogonal matrix.

The permutation matrix $\mathbf{M}$ is an orthogonal matrix.

%
%
%
%
%
%
%

Therefore, the eigenvalue of ${\mathbf{\Lambda }}^{n}$ (or $\mathbf{L}^{n}$) is not changed in the process of the orthogonal transformation to obtain $\mathbf{L}_{2}^{n}$, in essence. Thus ${{H}_{0.5}}\left( \mathbf{L}_{{}}^{n} \right)={{H}_{0.5}}\left( \mathbf{ML}_{{}}^{n}\mathbf{M} \right)$, the symmetry of system entropy is proved.

The symmetry of the system entropy function is proved, and from that we can draw more general inferences. For example, the serial numbers of nodes in the entire network are rotated, the value of the system entropy function remain the same, and the permutation matrix is as follow

\[\mathbf{ M } = \left[ \begin{matrix}
	0 & 1 & {} & {} & {} & {} & {}  \\
	{} & 0 & 1 & {} & {} & {} & {}  \\
	{} & {} & 0 & \ddots  & {} & {} & {}  \\
	{} & {} & {} & \ddots  & \ddots  & {} & {}  \\
	{} & {} & {} & {} & \ddots  & 1 & {}  \\
	{} & {} & {} & {} & {} & 0 & 1  \\
	1 & {} & {} & {} & {} & {} & 0  \\
\end{matrix} \right]\]

What's more, for any more general permutation, it can also get the corresponding orthogonal transformation that performs the permutation without changing the eigenvalues of the matrix ${\mathbf{\Lambda }}^{n}$. System entropy can be understood as a one-dimensional statistic of topology. Once the network topology is determined, the system entropy should be determined immediately, and the performance of network, such as power consumption and delay, should also be evaluated accurately. How we number the nodes should not affect the entropy of the final calculation. The above derivation and discussion can explain the rationality of system entropy function to describe network performance based on topology.

\subsection{non-negativity}

The non-negative property of system entropy indicates that system entropy also has the physical meaning as general entropy value. It is also an important evidence of the rationality of the system.

To prove this property, we need to prove $tr\left( {{\left( {{\mathbf{L}}^{n}} \right)}^{\frac{1}{2}}} \right)>1$.

According to the properties of the matrix, the trace of the analysis matrix is equivalent to the sum of the eigenvalues of the analysis matrix. Thus, let's look at the equivalent expression $tr\left( {{\left( {{\mathbf{\Lambda }}^{n}} \right)}^{\frac{1}{2}}} \right)$. First, we need to know the value of minimum eigenvalue of matrix  $  {{\mathbf{\Lambda }}^{n}}  $ is zero. Because the normalized Laplace matrix ${{\mathbf{L }}^{n}}$ is a semidefinite matrix. Since other eigenvalues of the normalized Laplace matrix can be understood as random variables in a certain range, it is difficult to directly analyze their sum.

So we need to analyze the range of eigenvalues of the normalized Laplace matrix, in particular, the maximum eigenvalue.

 In this case, the range of maximum eigenvalue of the normalized Laplace matrix is required. The value of maximum eigenvalue belongs to range $1<{{\lambda }_{N-1}}\le 2$. The upper bound "2" is not available for the normalized Laplace matrix of all graphs, because the nodes in such graphs are divided into two sets, the connecting structure is similar to the mapping of functions, and similar to the factor graph structure of LDPC, which belongs to bipartite graph.

However, the maximum eigenvalue of the normalized Laplace matrix of a nonbinary graph can be described as \cite{Complex Networks: A Networking and Signal Processing Perspective}

\begin{equation}
    \frac{N}{N-1}\le{{\lambda }_{N-1}} <2
    \label{EigRange}
\end{equation}

The result of takeing square root of anything greater than 1 is also greater than 1.So $1<\sqrt{{{\lambda }_{N-1}}}$ remains established. Thus $tr\left( {{\left( {{\mathbf{L}}^{n}} \right)}^{\frac{1}{2}}} \right)>1$ and ${{H}_{0.5}}\left( \mathbf{L}_{{}}^{n} \right)\ge 0$ are established, the nonnegativism of the system entropy function is proved.

\subsection{System Entropy in System of Etropy Theory}

This subsection will summarize the content of the previous three subsections and discuss whether the system entropy can be incorporated into a more generalized system of entropy theory.

In the first three subsections, the convexity, monotonicity, symmetry and nonnegativity of system entropy function are introduced and proved, which notice that it has convexity and monotone in terms of the variation of entropy value, and has non-negativity and symmetry in terms of the numerical stability of entropy. In other words, the properties mentioned above analyze the properties of system entropy from the perspective of change and constancy respectively. These properties can be related to changes in entropy caused by changes in topology and stability of entropy when topology is stable.

These properties of system entropy inevitably remind us of some properties of Shannon entropy, and we will compare these properties with the corresponding properties of Shannon entropy, thus drawing an important conclusion that it is reasonable to include system entropy in the entropy function system.

For example, shannon entropy system also has the property of symmetry. Shannon entropy is expressed as shannon entropy calculated for random vectors will not change because of the cyclic shift operation of random vectors. Corresponding to the system entropy proposed in this paper, both first-order statistics, entropy value will not be affected by the traversal order. It can be seen from this example that system entropy is also a member of the entropy function system.

In order to better compare Shannon entropy with system entropy, we compare the proven properties of system entropy with Shannon entropy, which are summarized in the following table:

\begin{table}[h!]
\centering

	\begin{center}

		\caption{Comparison of Shannon entropy and system entropy}
		\begin{tabular}{|p{2cm}<{\centering}|p{5cm}<{\centering}|p{5cm}<{\centering}|} 
			\hline
			{property} & system entropy & Shannon entropy \\
			\hline
			{convexity}  & \multicolumn{1}{p{5cm}}{\makecell[c]{${{H}_{0.5}}\left( \beta \mathbf{L}_{1}^{n}+\left( 1-\beta  \right)\mathbf{L}_{2}^{n} \right)\ge$  \\  $\beta {{H}_{0.5}}\left( \mathbf{L}_{1}^{n} \right)+\left( 1-\beta  \right){{H}_{0.5}}\left( \mathbf{L}_{2}^{n} \right)$}}  &
			\multicolumn{1}{|p{5cm}|}{\makecell[c]{$H\left( \theta {{\mathbf{p}}_{1}}+\left( 1-\theta  \right){{\mathbf{p}}_{2}} \right)\ge$ \\ $\theta H\left( {{\mathbf{p}}_{1}} \right)+\left( 1-\theta  \right)H\left( {{\mathbf{p}}_{2}} \right)$}} \\
			\hline
			symmetry  & ${{H}_{0.5}}\left( \mathbf{L}_{{}}^{n} \right)={{H}_{0.5}}\left( \mathbf{ML}_{{}}^{n}\mathbf{M} \right)$  & {\makecell[c]{$H\left( {{p}_{1}},{{p}_{2}},\cdots ,{{p}_{n}} \right)=$  \\  $H\left( {{p}_{{{j}_{1}}}},{{p}_{{{j}_{2}}}},\cdots ,{{p}_{{{j}_{n}}}} \right)$}} \\
			\hline
			nonnegativity  & ${{H}_{0.5}}\left( \mathbf{L}_{{}}^{n} \right)\ge 0$  & $H\left( \mathbf{p} \right)\ge 0$ \\
			\hline
		\end{tabular}
	\end{center}
\end{table}


As for other important properties of Shannon entropy, such as additivity and expansibility, expressions similar to additivity or expansibility need to be added in combination with specific application scenarios, that is, physical meanings of topological changes are closely related. The properties of system entropy will be an important research direction in the future.

\section{System entropy numerical results}

This section will introduce some numerical results of system entropy. In order to better explain the rationality of system entropy as a network topology evaluation criterion, this section presents four experiments to evaluate it. The first subsection presents the difference of the calculated results of system entropy for different types of network topologies. The second subsection presents some numerical results of convexity of system entropy. In the third subsection presents the corresponding relationship between system entropy and theoretical values of some practical networks. The fourth subsection presents the relationship between system entropy and average shortest path.

\subsection{Comparison of System entropy examples}

In this subsection, we have completed the simulation and analysis of two kinds of networks:  ER random network and NW small world network. The simulation methods of the networks are as follows:
\begin{itemize}
	\item [1)]
	For the generation of ER random network, network scale and edge connection probability need to be input.
	\item [2)]
	For NW small world network, average degree is given first and regular network is generated. Then the structure of the regular network is reconnected with broken edges without changing the number of edges in the network.
\end{itemize}
Our simulation strategy is to control the size of the network and the same number of edges of the network. This consistency requires the control of edge probability in ER random network, and the average degree in NW small world network, so that the topologies generated by these two networks become quantitatively consistent. Combined with the above two network generation methods, taking the node degree of 8 and the scale of 100 as an example, the simulation configuration needs to be set as follows:
\begin{itemize}
	\item [1)]
	ER random network: the number of nodes is set to 100, and the connection probability is set to 0.0808 $(100*8/2/ (100*99/2))$.
	\item [2)]
	NW small world network: the number of nodes is set to 100, and the average node degree is set to 8.
\end{itemize}
When simulating other node scales, the simulation configuration is carried out in a similar way.
\begin{figure}
	\centering
	\includegraphics[width=10cm]{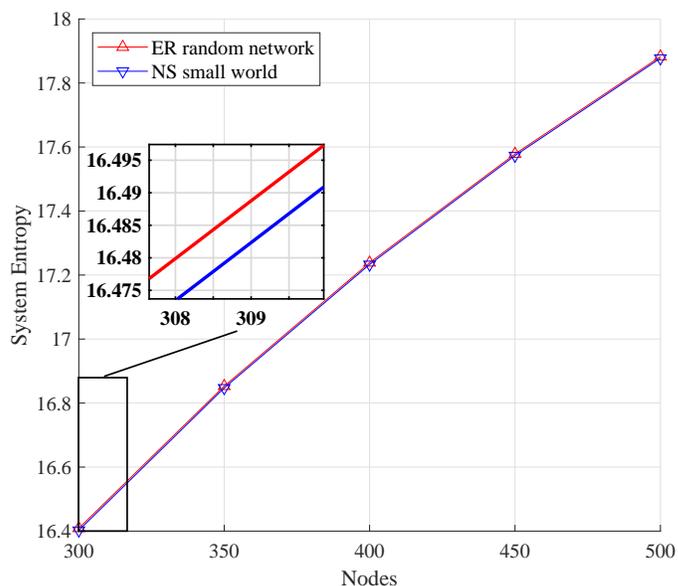}
	\caption{The system entropy varies with node size}
	\label{2}
\end{figure}
The reasons for such control are as follows: The system entropy, which measures network performance, is closely combined with topology. What we pay attention to is that the topology structure organized by the same resources of the network is reflected in the system entropy less. Therefore, in the process of simulation, each algorithm should be able to organize the same resources. Our simulation work is set up in the scene of increasing network scale to express the change of system entropy. Figure.\ref{2} shows the simulation results when the control average node degree is 8, and the network scale varies from 300 to 500.

It can be seen from the simulation results that the network performance reflected by system entropy is very close. However, if we pay attention to this slight difference, we can find that the system entropy of the small-world network is lower, i.e., NW small world network is the superior topology of the two network structures evaluated by system entropy. This implies that NW small world network can organize a network with better comprehensive performance under the premise of the same number of edges and nodes.

\subsection{Convex Property of System entropy examples}

\begin{figure}
	\centering
	\includegraphics[width=12cm]{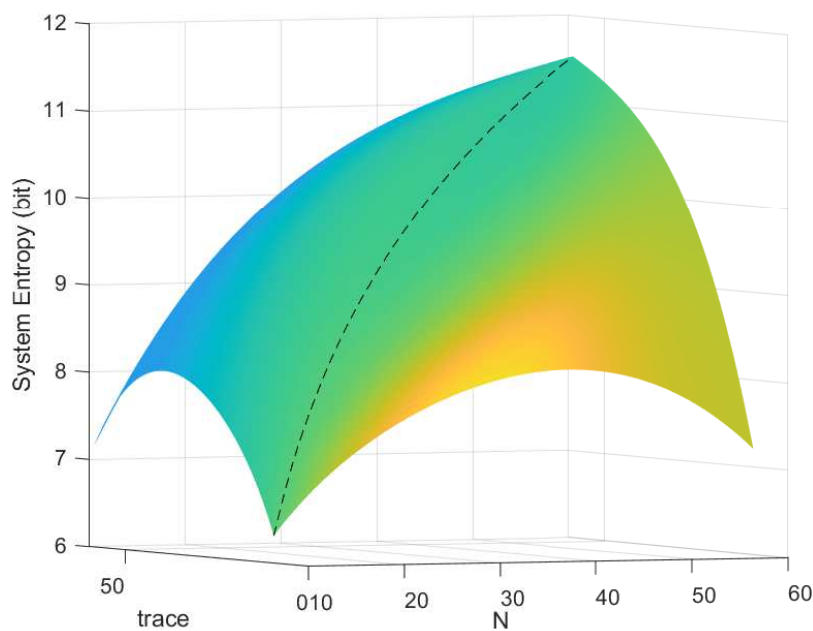}
	\caption{The convexity of system entropy function}
	\label{3}
\end{figure}

This subsection gives some numerical results that show convexity of system entropy. The experiment in this subsection is carried out according to the following ideas: the trace of the Laplace matrix is directly related to the entropy value of the system entropy. However, it is not easy to directly control the trace of the Laplace matrix. Therefore, we obtain a large number of samples by generating a large number of Laplace matrices, and statistic these samples to obtain certain rules. The results were presented using three-dimensional coordinates as figure.\ref{3}, so that the convexity of system entropy is discussed in terms of node scale and sum of eigenvalues of taken squared Laplace matrix.

 Figure.\ref{3} shows that the entropy of the system shows convexity with the increase of node size and sum of eigenvalues. Therefore, the experiments in this subsection show that the entropy function of the system is numerically convex.

\subsection{The System Entropy of the Real Network}

In order to prove the rationality of the evaluation of the system entropy function on the industrial network, this section will give the value of the system entropy in the example of the real network, so as to explain the rationality of the system entropy index.

The network in the example analyzed was generated from E-mail data from a large European research institution. The open source dataset has been anonymized for the specific data in the network, and now only the topology of the network can be seen. The degree of the network is about 16. At the same time, the network is divided into four main groups, the number of nodes are 309, 162, 89 and 142 respectively. These four groups are all small-world networks. In order to intuitively express the relationship between the system entropy of these four actual networks and the system entropy curve obtained by computer simulation, we plotted the four discrete points together with the corresponding curve to observe the changes.
\begin{figure}
	\centering
	\includegraphics[width=10cm]{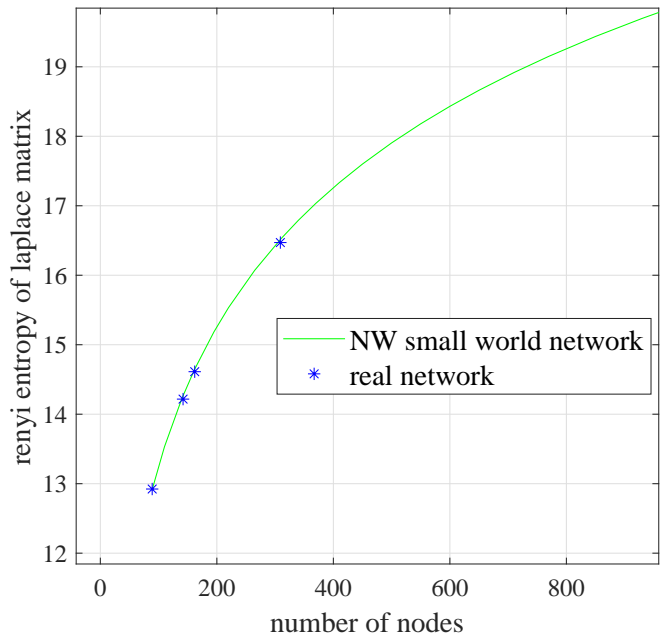}
	\caption{system entropy of real and simulated NW small world network}
	\label{realNet}
\end{figure}

Figure.\ref{realNet} shows the simulation results under the above conditions. It can be seen from the simulation results that the system entropy values of the four networks are distributed near the curve of degree 16, and the fitting degree is very high. This result indicates that the system entropy values of the small-world network instance are very close to those of the simulated small-world network.

\subsection{The Relationship between APL and System Entropy}

\begin{figure}
	\centering
	\includegraphics[width=10cm]{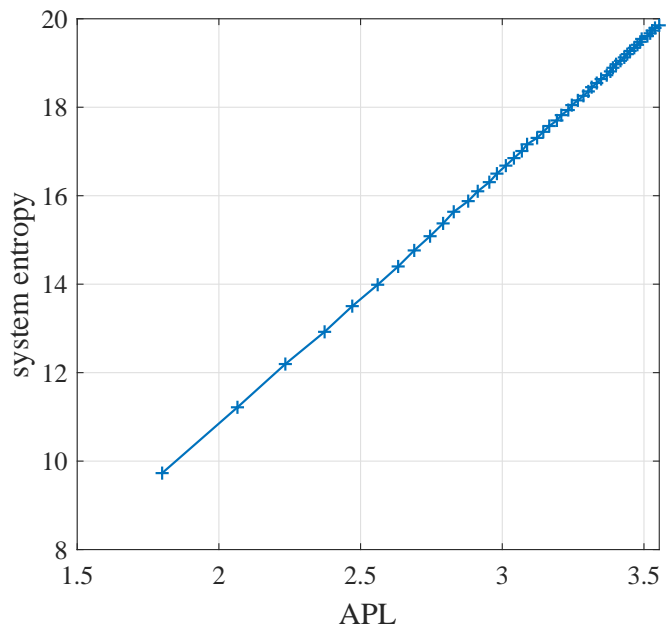}
	\caption{APL vs system entropy of NW small world network}
	\label{aplvsSE}
\end{figure}

In order to further verify the rationality of the system entropy function, we combined the APL to give the simulation results. The reason why the APL is used as the reference of system entropy is that the APL can describe the number of nodes required for the average information transmission in the network, and is an important indicator of the information transmission speed in the network. According to the corresponding relationship above, the shorter the APL is, the lower the system entropy of the network should be, that is, the shorter APL brings higher transmission performance. The simulation configuration is as follows: in this experiment, NW small-world network is used to generate a network model, and the average minimum distance of the generated network and the system entropy value of the network topology are counted. If the APL is used as the horizontal coordinate, the system entropy of the network is used as the vertical coordinate the simulation results are as figure.\ref{aplvsSE}.

By drawing the APL and system entropy in a graph, it can be seen that: the system entropy function of the network is proportional to the APL; This phenomenon can be explained as follows: The longer the APL, the larger the entropy function value of the system, the more disordered the network, and the longer the APL and longer delay are required for the transmission of information. The linear proportional relationship between APL and system entropy in this experiment shows that system entropy can realize the evaluation of network power consumption in the sense of measuring the forwarding times and data transmission distance of the network, which further proves that system entropy has great guiding significance for measuring network performance.

\section{Conclusion and perspective}
This paper explores network models applied in industrial networks from a network scale perspective. This paper summarizes the key definitions and metrics of graph theory related to network models, and focuses on three network models: random network, small-world network, and scale-free network. On the basis of the above models, this paper continues to discuss the network in the existing industrial applications, and establishes the mapping relationship between the industrial application and the network model. Finally, this paper proposes a novel network performance metric-system entropy, and analyzes the mathematical properties of this metric. The system entropy can well cover a variety of network models and achieve a unified performance comparison under different models. It is shown that the system entropy has the potential for further expansion and research.

\Acknowledgements{This work was supported by the Key Program of National Natural Science Foundation of China (No. 92067202), National Natural Science Foundation of China (No. 62071058) and the Key Laboratory of Universal Wireless Communications (BUPT), Ministry of Education, P.R.China (No.KFKT-2022104).}




\bibliographystyle{unsrt}
\end{document}